# The communication of meaning in social systems

Loet Leydesdorff [*] & Sander Franse

**Abstract**

The sociological domain is different from the psychological one insofar as meaning can be communicated at the supra-individual level (Schütz, 1932; Luhmann, 1984). The computation of anticipatory systems enables us to distinguish between these domains in terms of weakly and strongly anticipatory systems with a structural coupling between them (Maturana, 1978). Anticipatory systems have been defined as systems which entertain models of themselves (Rosen, 1985). The model provides meaning to the modeled system from the perspective of hindsight, that is, by advancing along the time axis towards possible future states. Strongly anticipatory systems construct their own future states (Dubois, 1998a and b). The dynamics of weak and strong anticipations can be simulated as incursion and hyper-incursion, respectively. Hyper-incursion generates "horizons of meaning" (Husserl, 1929) among which choices have to be made by incursive agency.

[*] Amsterdam School of Communications Research (ASCoR), University of Amsterdam, Kloveniersburgwal 48, 1012 CX Amsterdam, The Netherlands, Email: loet@leydesdorff.net; Tel.: +31-20-525 6598; Fax: + 31-84223911.

## Introduction

In his mathematical theory of communication, Shannon detached himself from the communication of meaning by stating on the first page that the “semantic aspects of communication are irrelevant to the engineering problem” (Shannon & Weaver 1949, at p. 3). However, his co-author Weaver noted the following:

> The concept of information developed in this theory at first seems disappointing and bizarre--disappointing because it has nothing to do with the meaning, and bizarre because it deals not with a single message but rather with the statistical character of a whole ensemble of messages, bizarre also because in these statistical terms the two words *information* and *uncertainty* find themselves to be partners.
>
> I think, however, that these should be only temporary reactions; and that one should say, at the end, that this analysis has so penetratingly cleared the air that one is now, perhaps for the first time, ready for a real theory of meaning. (*Ibid.*, at pp. 116f.)

Recent advances in the computation of anticipatory systems enable us to model both the generation of meaning by anticipatory agents and meaning-processing at the supra-individual level.

Luhmann (1971, 1984, and 1986) proposed to consider the processing of meaning as the *autopoietic* operation of both social and psychological systems. Social and psychological systems are both structurally coupled and “interpenetrate” each other reflexively (Luhmann, 1977). Interpenetration adds an operational dimension to the structural coupling between social and psychological systems (Luhmann, 1988 [1995a, at p. 51; 2002, at p. 182]). In this study, we submit operationalizations of these various concepts in terms of the theory and computation of anticipatory systems (Rosen, 1985; Dubois, 1998a; Leydesdorff, 2008, 2009).

Meaning can first be provided to events by reflexive systems (e.g., observers) from the perspective of hindsight (Rosen, 1985). Human languages allow additionally for the construction and exchange of models using metaphors (Luhmann, 1995a, at p. 44; 2002, at p. 175; Leydesdorff & Hellsten, 2005). Metaphors enable us to communicate meaning (Lakoff & Johnson, 1980; Hesse, 1988; Maasen & Weingart, 1995). New meaning can also be generated as a result of inter-human communication (Schütz, 1932; Mead, 1934). When expectations are exchanged and interact, one can expect the development of a non-linear dynamics of meaning-processing on top of the information exchanges (MacKay, 1969; Maturana, 2000). Meaning is interactively and recursively reconstructed, but in an intentional mode, i.e., with reference to a future state (Husserl, 1929; Schutz, 1975).

The perspective of hindsight—that is, entertaining a model which provides meaning to the modeled system (Rosen, 1985)—can itself be modeled as the backward evaluation of a difference equation (Dubois, 1998a). This has been called “incursion” in order to distinguish it from “recursion” which follows the arrow of time. For example, the well-known logistic map $x_{t+1} = ax_t\ (1 - x_t)$ can *incursively* be formulated as follows:

$$x_{t+1} = ax_t(1 - x_{t+1}) \quad (1)$$

and

$$x_{t+1} = ax_{t+1}(1 - x_t) \tag{2}$$

In the case of Equation 1, the selection pressure $(1 - x)$ on the variation $(x)$ develops synchronously with the recursive development of the system. For example, markets select technological innovations in terms of current prices, while competing technologies can be expected to develop with reference to their previous states. Equation 2 has two roots: $x = (a - 1)/a$, and $x = 0$. These roots are the steady states of the logistic equation and the models of anticipatory systems that can be derived from it.

By considering the future states as the drivers of the system in the present, three *hyper-incursive* equations can additionally be formulated:

$$x_t = ax_t(1 - x_{t+1}) \tag{3}$$

$$x_t = ax_{t+1}(1 - x_t) \tag{4}$$

$$x_t = ax_{t+1}(1 - x_{t+1}) \tag{5}$$

These equations model a "strongly anticipatory" system. While "weakly anticipatory" systems entertain models of themselves, strongly anticipatory ones use expectations to construct their current state. In other words, the incursive and hyper-incursive analogues of the logistic equation provide us with a model of how social systems of expectations can be reconstructed on the basis of interactions among weakly anticipatory systems (Leydesdorff & Dubois, 2004).

## The incursive model

Using Equation 1, one can simulate a modeling system (Leydesdorff, 2005). One can also consider the modeling system as an observer (Spencer-Brown, 1969). The model appreciates the modeled system by filtering the noise, as can be understood by rewriting as follows:

$$x_{t+1} = ax_t(1 - x_{t+1}) \tag{1}$$

$$x_{t+1} = ax_t - ax_t x_{t+1} \tag{1a}$$

$$x_{t+1}(1 + ax_t) = ax_t \tag{1b}$$

$$x_{t+1} = ax_t /(1 + ax_t) \tag{1c}$$

Unlike the logistic map, this anticipatory system does not bifurcate, but develops along the curve of the steady state—$x = (a - 1)/a$—for all values of $a$. Figure 1 shows the results of the simulation. The biological variation bifurcates and increasingly generates chaos for $3.57 < a < 4$, while the anticipatory system grows continuously to the limit value of $x = 1$ with increasing values of $a$.

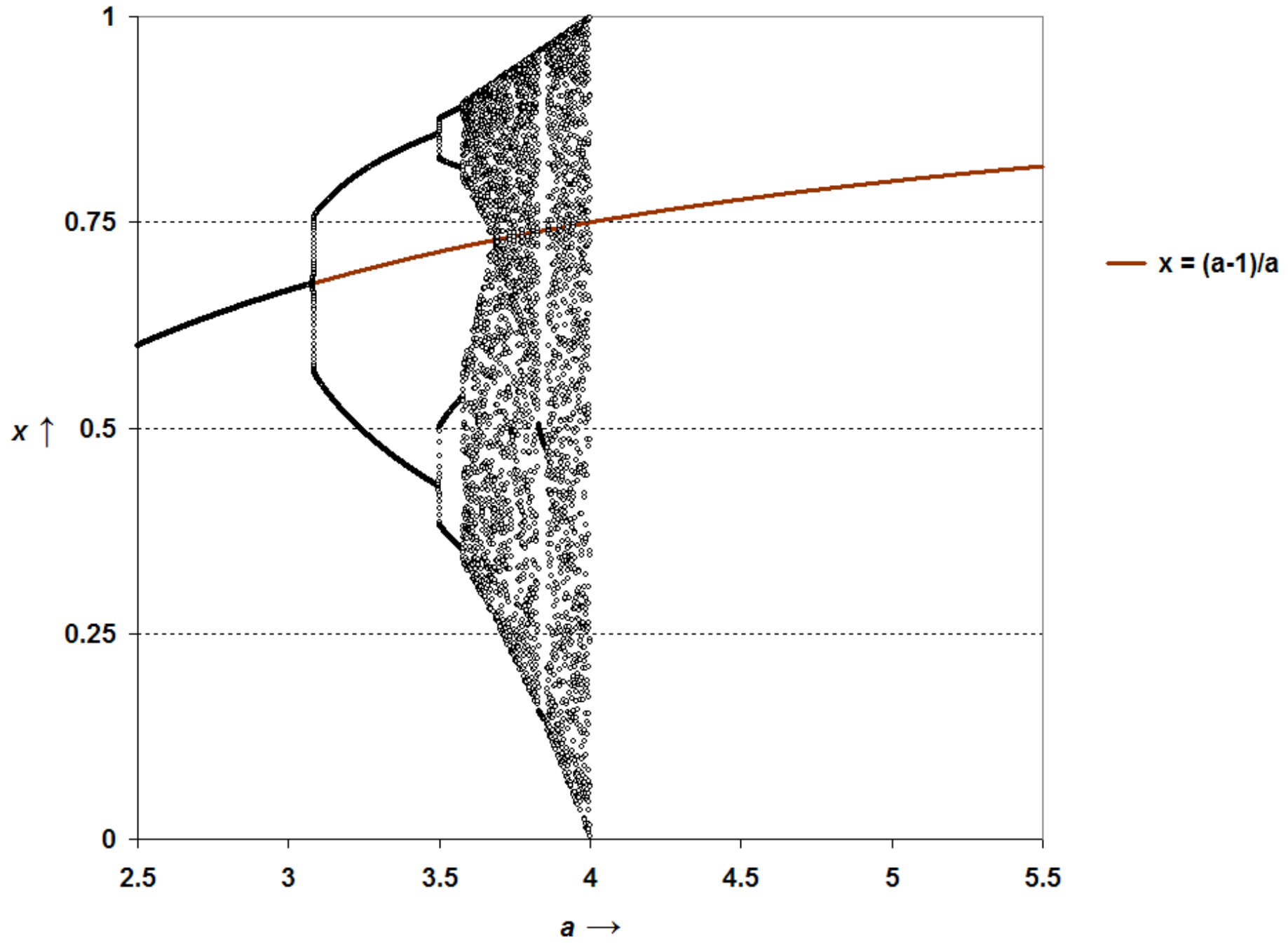

**Figure 1**: The steady state of the weakly anticipatory system.

The line penciled into Figure 1 can be considered as an emerging axis stabilizing an identity among the reflections at each moment of time. The weakly anticipatory system provides meaning to the events by integrating them in both the biological domain ($a < 4$; e.g., bodily perceptions) and the social domain of meaning-processing ($a \geq 4$). This integration of the different representations functions as a linchpin for developing a strongly anticipatory system in the cultural (i.e., non-natural) domain of meaning-processing ($a \geq 0.4$).

## The hyper-incursive model

The hyper-incursive Equations 3, 4, and 5 provide us with the three building blocks of a non-linear dynamics of expectations (May, 1976; May & Leonard, 1975). Equation 3 first evolves into $x = (a - 1)/a$. This is equal to the steady state of the weakly anticipatory system. In other words, weak anticipation can be considered as one of the sub-dynamics of a strongly anticipatory system. In Luhmann's terminology, one could perhaps consider this correspondence as the re-entry of the first-order observation into the system of second-order observations (Luhmann, 1993 [1999]).

Equation 4 evolves into $x_{t+1} = (1/a)\ [x_t / (1 - x_t)]$. This routine formalizes the reflexive operation: when $x_t > [a / (1 + a)]$ a pulse is generated which first overshoots the value of one,[1] but then generates a negative value (Figure 2). The negative value provides a mirror image of the representation at a specific moment in time, and thus allows for a reflection. Reflection enables a system to bounce a communication. Note that the combination of Equations 3 and 4 provides the weakly anticipatory system with reflexive access to the meaning processing at the supra-individual level.

---

[1] $(1/a)[x_t / (1 - x_t)] = 1$ for $x_t = a / (1 + a)$.

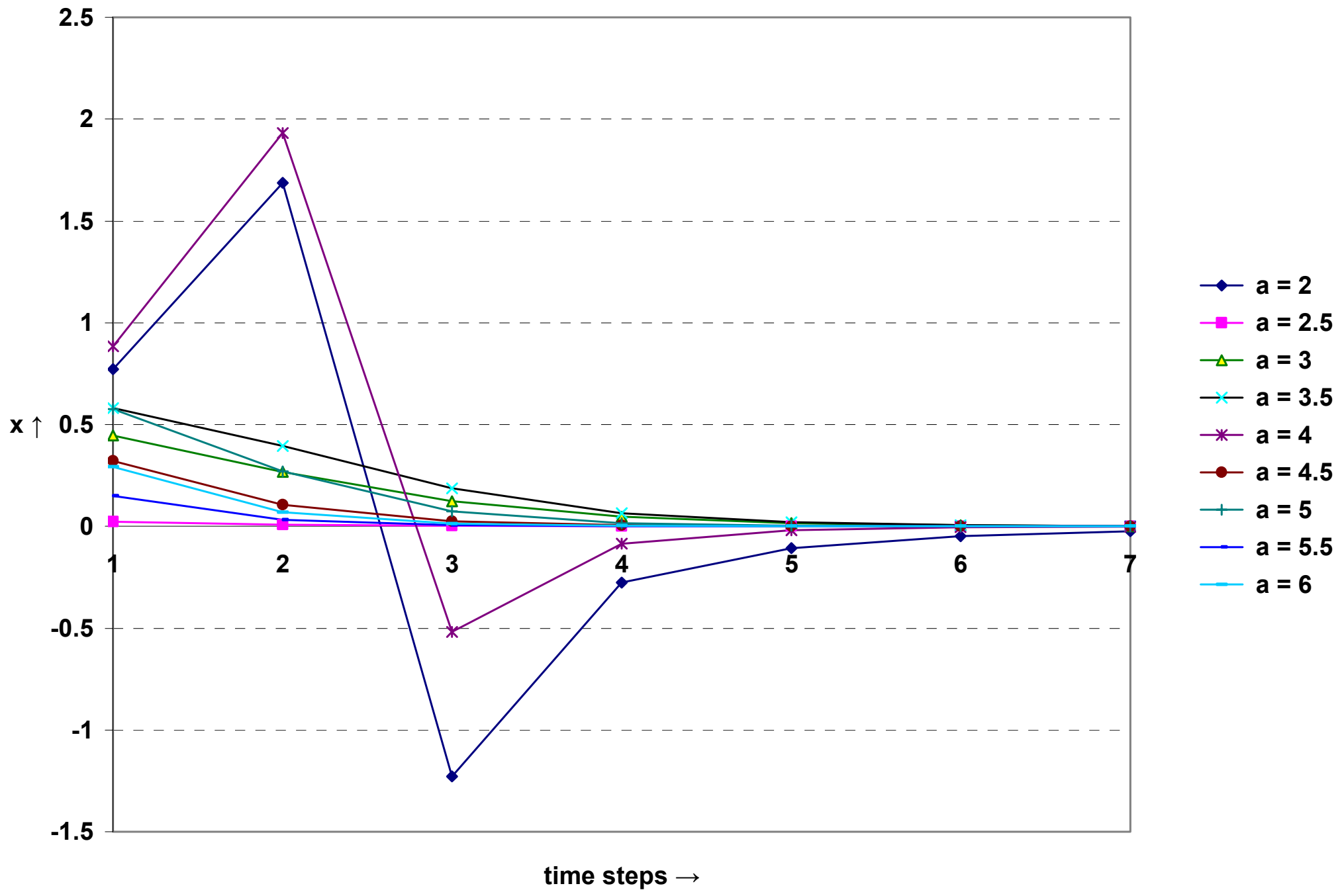

**Figure 2:** Simulation of Equation 4.

Equation 5 invokes the perspective on social systems as introduced here: *Ego* and *Alter* are bound by a *double* contingency of expecting each other to be first historical and biological, but secondly reflexive and intentional (Parsons, 1968; Luhmann, 1984). Using the second layer of the double contingency, *Ego* in the present ($x_t$) no longer refers to oneself as an identity which is rooted in the past, but to oneself in a future state ($x_{t+1}$), that is, as an *Alter Ego*. The non-linear interactions among expectations can generate the social world as a system different from psychological ones. However, this social system of expectations remains structurally coupled to psychological ones; otherwise, nobody would be able to articulate the "horizons of meaning" which are generated.

Social coordination mechanisms are only perceptible by human beings after a reflexive turn (Giddens, 1984). Let us now show that reflexive decision-making is endogenous to the social system of expectations (Dubois, 1998b). Equation 5 can be rewritten as follows:

$$x_t = ax_{t+1}(1 - x_{t+1}) \tag{5}$$

$$x_t = ax_{t+1} - ax_{t+1}^2$$

$$ax_{t+1}^2 - ax_{t+1} + x_t = 0$$

$$x_{t+1}^2 - x_{t+1} + x_t / a = 0$$

In general, this equation has two solutions:

$$x_{t+1} = ½ \pm ½ \sqrt{[1 - (4/a)\, x_t]} \tag{6}$$

Given that $0 \leq x \leq 1$, the curve $x = 0.5 \pm 0.5 \sqrt{(1 - (4/a))}$ for $x = 1$ sets limits to the possible values reached by the social system (Figure 3).

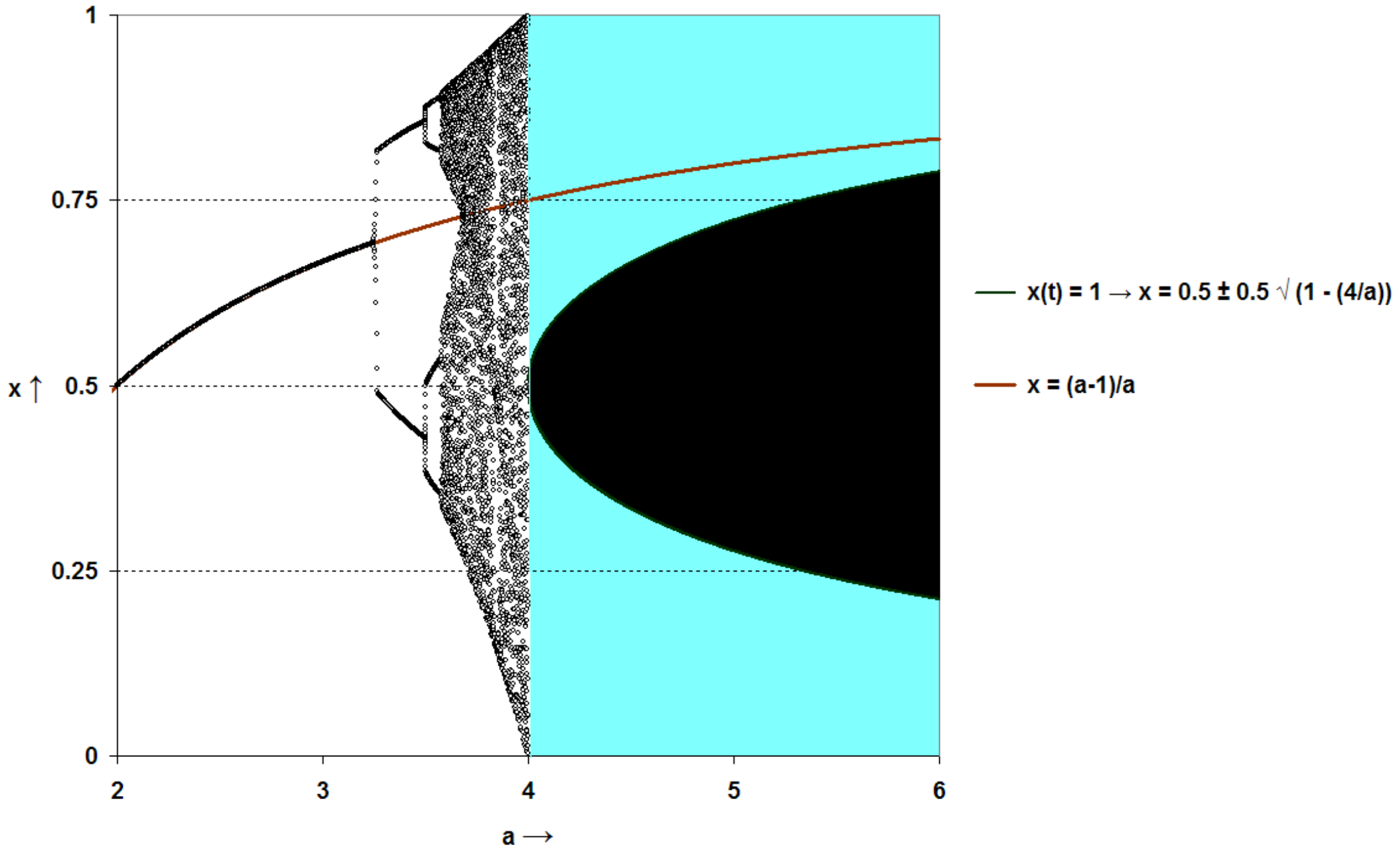

**Figure 3**: The social system as a result of hyper-incursion.

For $a \geq 4$, two sets of expectations are generated at each time step depending on the plus or the minus sign in the equation. After $N$ time steps, $2^N$ future states would be possible. Thus, the social system of expectations needs continuously a mechanism for making decisions between options because otherwise this system would rapidly become overburdened with uncertainty.

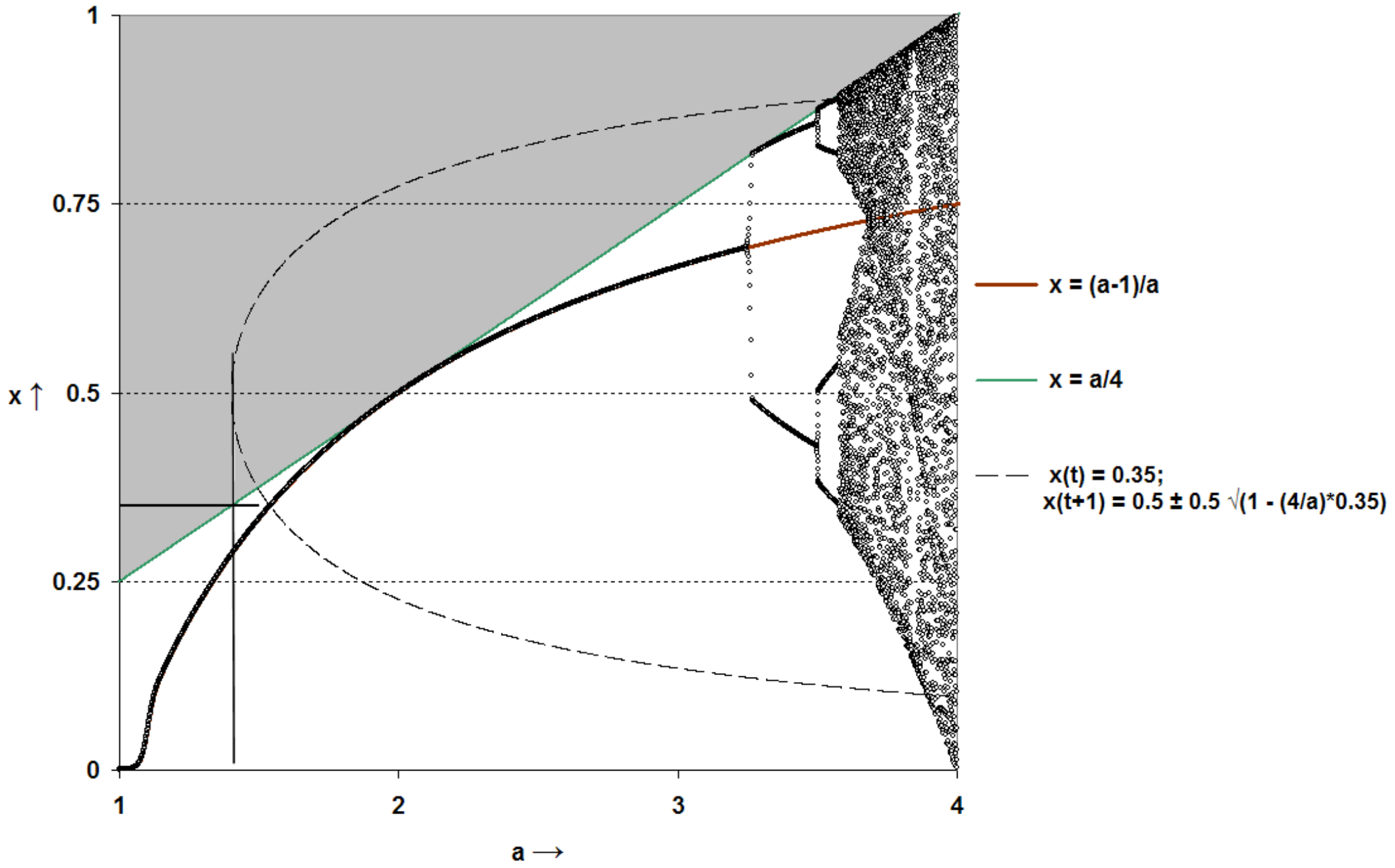

**Figure 4**: Possible penetrations of the social system into the biological variation ($a < 4$).

The term under the root in Equation 6 is positive for $x_t \leq a/4$: this condition is met for $a \geq 4$, but sets a borderline to the possible penetrations of the social system into the biological variation ($a < 4$). In Figure 4, this limitation is elaborated for $a = 1.2$ and $x_t = 0.35$. Since in a next step $x_{t+1} = 0.5$, and thus larger than $a/4$ (= 0.3), the strongly anticipatory system would not be able to proceed with a next step to $x_{t+2}$. Expectations cannot operate on expectations for $a < 4$, but next states of the natural system can be constructed on the basis of expectations like in the case of the socio-economic construction of technologies (Leydesdorff, 2006).

### Decisions and historical trajectories

The social system cannot further be developed at $a \geq 4$ without a form of agency taking decisions because of the continuous production of uncertainty by the hyper-incursive mechanism (Eq. 5). Luhmann (2000) identified decisions as structuring organizations. Although this reflexive capacity is conceptualized by Luhmann as endogenous to the social system, organizations can also be attributed with institutional agency. A psychological system can then perhaps be considered as the minimal unit of reflection for making choices (Habermas 1981; Leydesdorff 2000, 2001). If decisions are socially further organized—for example, by using decision rules—an institutional layer can increasingly be shaped. The institutional layer provides a retention mechanism for the next round of expectations (Aoki 2001; Luhmann 2000). Thus, the social system is dually layered as a forward-moving retention mechanism and sets of possible expectations which flow through the networks. These "horizons of meaning" are not given, but continuously undergoing reconstruction (Luhmann 1990, 2002).

The strongly anticipatory system is not autonomous, but *autopoietic* since structurally coupled to the layer of decision-making by weakly anticipatory agents (Collier 2006). Because the equations are derived from the same logistic equation, the steady states of the systems are equal: the systems are not only structurally coupled, but also coupled in terms of their reflexive operation. As noted, Luhmann (1977, 1995a [2002]) reserved the term "interpenetration" (Parsons, 1968, at p. 473; Luhmann, 1978) for this additional dimension in the coupling. Interpenetration is possible because of the evolutionary achievement of developing human language ((Luhmann, 1995a, at p. 44; 2002, at p. 175; Leydesdorff 2002). Communication provides the strongly anticipatory system with one degree of freedom more than the weakly anticipatory ones. Unlike *in*dividuals, the social system can be expected to remain distributed. The uncertainty in the double contingency (Equation 5) is reproduced by the system and this makes the hyper-incursive anticipations unstable flows of communication.

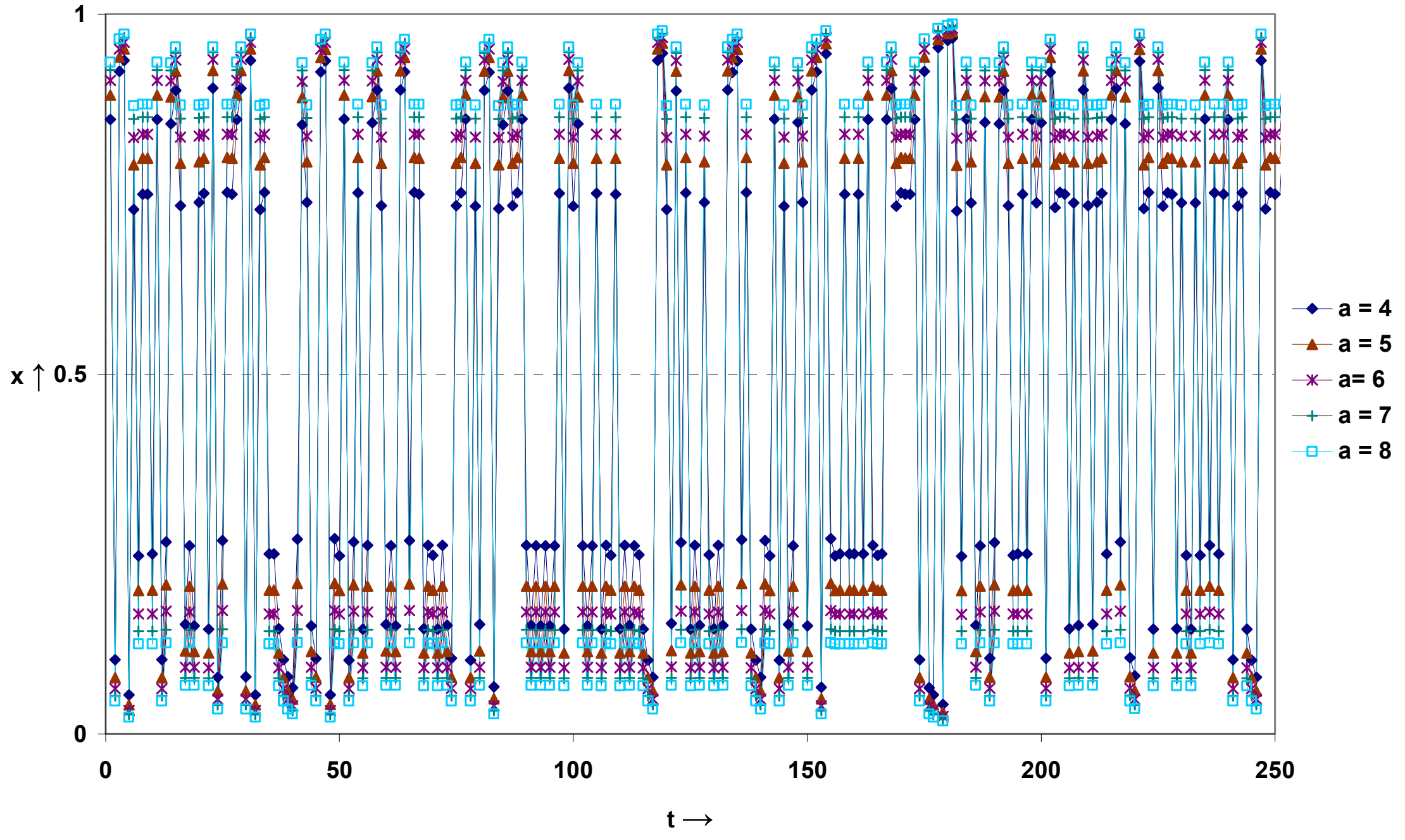

**Figure 4**: Trajectory of a social system based on random decisions of the decision-making units, for various values of the bifurcation parameter *a*.

Figure 4 shows a simulation of trajectories for different values of *a* where decisions are taken randomly. The system sometimes dwells in a specific state. Next-order mechanisms like institutionalization can be expected to stabilize these configurations because decisions are then no longer taken randomly (Dubois 2001).

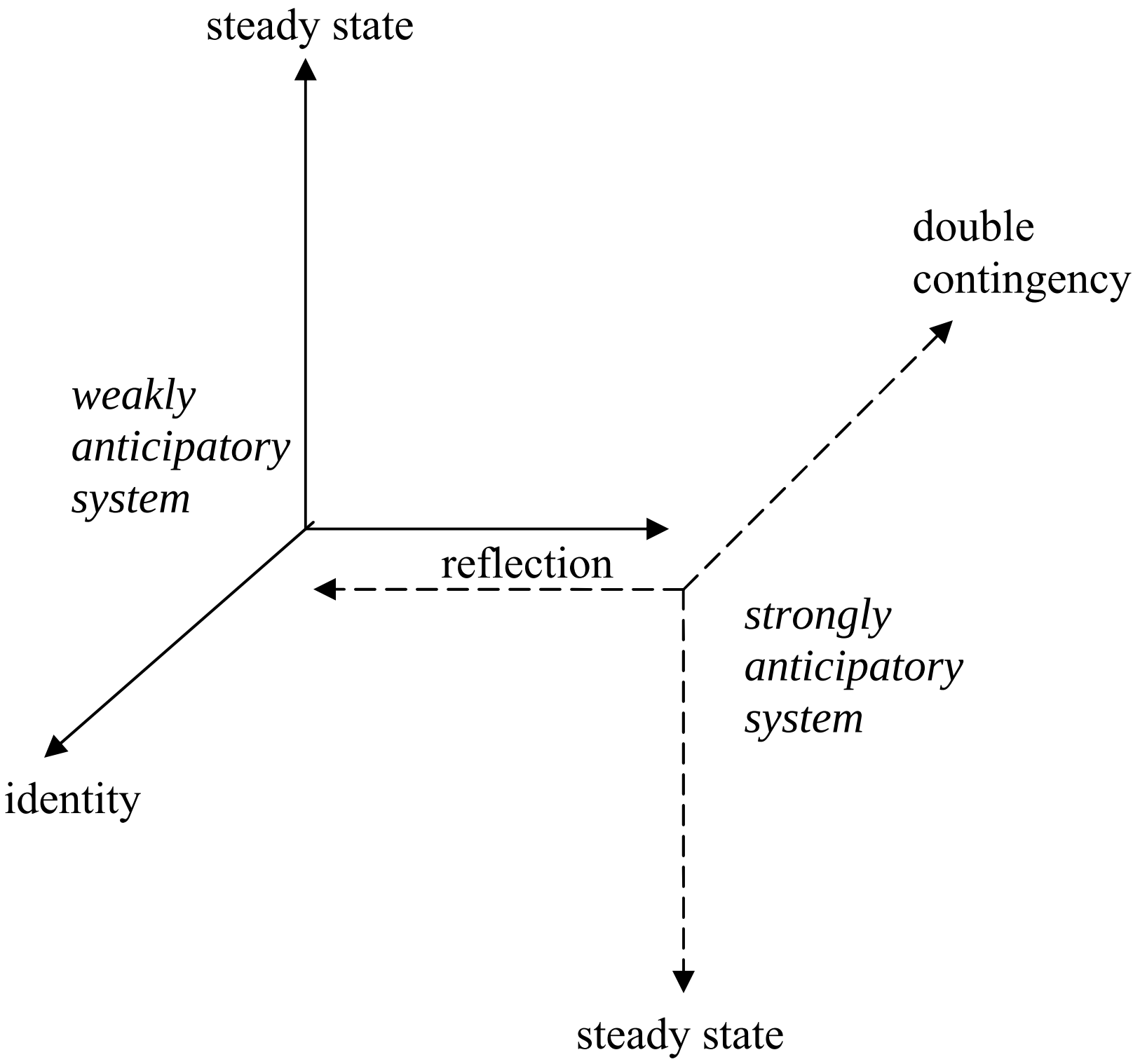

**Figure 5:** The relations among the various subdynamics

Figure 5 summarizes the subdynamics which were distinguished above. While the weakly anticipatory system tends to integrate representations by organizing them into an identity, the strongly anticipatory one is based on uncertainty contained in the distribution. However, this additional degree of freedom cannot be used by this system without the mediation of agency using the additional coupling which is provided reflexively. The two systems are not only structurally coupled as systems, but also in terms of the reflexive operation of providing meaning to each other's operations.

## Conclusions

The incursive and hyper-incursive equations provide us with models of meaning-processing at the individual and supra-individual levels. The incursive equation models the operation of a modeling system in relation to the modeled one. This corresponds with Rosen's (1985) definition of an anticipatory system. The hyper-incursive equations enable us to model the structural coupling between historical phenomena and emerging horizons of meaning. Since hyper-incursion and incursion can recursively be applied to the results of the operations, the modeling and simulation of further codification of meanings (e.g., into discursive knowledge) becomes feasible.

At the theoretical level, these algorithmic results can be appreciated with reference to the Habermas-Luhmann discussion (Habermas & Luhmann 1971) about communicative competencies (of agency) *versus* communication systems. The opposing perspectives in this debate can be considered as providing two geometrical metaphors to the algorithmic operations of the social system (Leydesdorff, 2000). Geometrical metaphors generate their respective "blind spots." Habermas (1981), for example, shares with Giddens (1984) the assumption that the operation of the social system of expectations remains necessarily "virtual" and therefore unspecifiable. Luhmann (1984) specified this operation theoretically by proposing a deliberate abstraction from human agency (Luhmann, 1995b).

Luhmann's theory could be used fruitfully as a heuristics in these simulations. However, our results suggest that because of the complex relation between structural coupling between the two types of systems, *and* the interpenetrating reflection of their mutual operations, decision-making and therefore agency can be considered as back on stage. This is endogenous to the system and can also be appreciated as a re-entry. In Luhmann's theory, the social system is considered as operationally closed and this organizational reflection can only be formulated as re-entry. However, the alternative perspective of considering the system as semi-autopoietic (since dependent for its further development on human or organizational reflections and agency) remains also possible. An advantage of this latter perspective may be that it is more compatible with mainstream sociological theorizing by appreciating reflexive agency while keeping the surplus value of Luhmann's theory.